\documentclass[submission]{eptcs}

\title{Sessions as Propositions}
\def\titlerunning{Sessions as Propositions}
\def\authorrunning{Lindley \& Morris}

\author{
  Sam Lindley
  \qquad \qquad
  J. Garrett Morris
\institute{The University of Edinburgh \\
  \email{\{Sam.Lindley,Garrett.Morris\}@ed.ac.uk}}}

\def\event{PLACES 2014}

\usepackage{amssymb}
\usepackage{amsmath}
\usepackage{color}
\usepackage{stmaryrd}
\usepackage{mathpartir}
\usepackage{xspace}

\newtheorem{theorem}{Theorem}

\newcommand{\ba}{\begin{array}}
\newcommand{\ea}{\end{array}}

\newcommand{\bl}{\ba{@{}c@{}}}
\newcommand{\el}{\ea}

\newenvironment{equations}{\[\ba{@{}r@{~}c@{~}l@{}}}{\ea\]}
\newenvironment{eqs}{\ba{@{}r@{~}c@{~}l@{}}}{\ea}

\newcommand{\key}{\mathsf}

\newcommand{\set}[1]{\{ #1 \}}

\newcommand{\cptogv}[1]{\llparenthesis{#1}\rrparenthesis}

\newcommand{\row}[2]{\set{#1}_{#2}}

\newcommand{\gvOutput}[2]{\mathord{!}{#1}.{#2}}
\newcommand{\gvInput}[2]{\mathord{?}{#1}.{#2}}
\newcommand{\gvEndOutput}{\key{end}_!}
\newcommand{\gvEndInput}{\key{end}_?}
\newcommand{\gvPlus}[2]{\oplus \row{#1}{#2}}
\newcommand{\gvChoice}[2]{\binampersand \row{#1}{#2}}
\newcommand{\gvServer}[1]{\flat {#1}}
\newcommand{\gvService}[1]{\sharp {#1}}
\newcommand{\gvDual}[1]{\overline{#1}}
\newcommand{\gvOutputType}[2]{\mathord{!}[{#1}].{#2}}
\newcommand{\gvInputType}[2]{\mathord{?}[{#1}].{#2}}

\newcommand{\cpj}[2]{{#1} \vdash {#2}}
\newcommand{\gvj}[3]{{#1} \vdash {#2} : {#3}}

\newcommand{\la}{l}
\newcommand{\G}{\Gamma}
\newcommand{\D}{\Delta}

\newcommand{\lolli}{\multimap}

\newcommand{\gvLinFun}[2]{{#1} \lolli {#2}}
\newcommand{\gvUnFun}[2]{{#1} \to {#2}}
\newcommand{\gvTimes}[2]{{#1} \otimes {#2}}

\newcommand{\gvLink}[2]{\key{link}~{#1}~{#2}}
\newcommand{\gvLam}[2]{\lambda {#1}.{#2}}
\newcommand{\gvApp}[2]{{#1}~{#2}}
\newcommand{\gvPair}[2]{({#1},{#2})}
\newcommand{\gvLet}[3]{\key{let}~{#1}={#2}~\key{in}~{#3}}
\newcommand{\gvSend}[2]{\key{send}~{#1}~{#2}}
\newcommand{\gvReceive}[1]{\key{receive}~{#1}}
\newcommand{\gvSelect}[2]{\key{select}~{#1}~{#2}}
\newcommand{\gvCase}[2]{\key{case}~{#1}~\key{of}~{#2}}

\newcommand{\gvFork}[2]{\key{fork}~{#1}.{#2}}

\newcommand{\gvWith}[3]{\key{with}~{#1}~\key{connect}~{#2}~\key{to}~{#3}}

\newcommand{\gvReceiveK}[4]{\gvLet{\gvPair{#1}{#2}}{\gvReceive{#3}}{#4}}

\newcommand{\gvSendType}[2]{\key{sendType}~{#1}~{#2}}
\newcommand{\gvReceiveType}[2]{\key{receiveType}~{#1}.{#2}}

\newcommand{\gvServe}[2]{\key{serve}~{#1}.{#2}}
\newcommand{\gvRequest}[1]{\key{request}~{#1}}

\newcommand{\cpLink}[2]{{#1} \leftrightarrow {#2}}
\newcommand{\cpCut}[3]{\nu {#1}.({#2} \mid {#3})}
\newcommand{\cpOutput}[4]{{#1}[{#2}].({#3} \mid {#4})}
\newcommand{\cpInput}[3]{{#1}({#2}).{#3}}
\newcommand{\cpInject}[3]{{#1}[{#2}].{#3}}
\newcommand{\cpCase}[2]{{#1}.\key{case}~{#2}}
\newcommand{\cpServe}[3]{!{#1}({#2}).{#3}}
\newcommand{\cpRequest}[3]{?{#1}[{#2}].{#3}}
\newcommand{\cpEmptyOut}[1]{{#1}[].0}
\newcommand{\cpEmptyIn}[2]{{#1}().{#2}}
\newcommand{\cpSendType}[3]{{#1}[{#2}].{#3}}
\newcommand{\cpReceiveType}[3]{{#1}({#2}).{#3}}

\newcommand{\cpTimes}[2]{{#1} \otimes {#2}}
\newcommand{\cpPar}[2]{{#1} \mathbin{\bindnasrepma} {#2}}
\newcommand{\cpPlus}[2]{\oplus \row{#1}{#2}}
\newcommand{\cpWith}[2]{\binampersand \row{#1}{#2}}
\newcommand{\cpOne}{1}
\newcommand{\cpBottom}{\bot}
\newcommand{\cpOfCourse}[1]{\mathord{!}{#1}}
\newcommand{\cpWhyNot}[1]{\mathord{?}{#1}}
\newcommand{\cpDual}[1]{{#1}^\bot}
\newcommand{\cpExists}[2]{\exists {#1}.{#2}}
\newcommand{\cpForall}[2]{\forall {#1}.{#2}}

\newcommand{\un}[1]{\mathit{un}(#1)}
\newcommand{\lin}[1]{\mathit{lin}(#1)}

\newcommand{\FV}[1]{\mathit{FV}(#1)}
\newcommand{\subst}[3]{{#1}[{#2}/{#3}]}

\newcommand{\last}{LAST\xspace}

\newcommand{\hgv}{HGV\xspace}
\newcommand{\hgvpi}{HGV$\pi$\xspace}

\newcommand{\lampi}[1]{({#1})^\star}
\newcommand{\hgvcp}[1]{\llbracket{#1}\rrbracket}

\newcommand{\hgvcpl}{\left\llbracket}
\newcommand{\hgvcpr}{\right\rrbracket}

\newcommand{\redto}{\longrightarrow}
\newcommand{\eqto}{\equiv}

\newcommand{\infr}[3][]{\inferrule{#2}{#3}}

\begin{document}

\maketitle

\begin{abstract}
  Recently, Wadler presented a continuation-passing translation from a session-typed functional
  language, GV, to a process calculus based on classical linear logic, CP. However, this translation
  is one-way: CP is more expressive than GV. We propose an extension of GV, called \hgv, and give
  translations showing that it is as expressive as CP. The new translations shed light both on the
  original translation from GV to CP, and on the limitations in expressiveness of GV.
\end{abstract}

\section{Introduction}
\label{sect:introduction}

Linear logic has long been regarded as a potential typing discipline for concurrency.
Girard~\cite{Girard87} observes that the connectives of linear logic can be interpreted as parallel
computation. Abramsky~\cite{Abramsky92} and Bellin and Scott~\cite{BellinScott94} interpret linear
logic proofs as $\pi$-calculus processes.  While they provide $\pi$-calculus interpretations of all
linear logic proofs, they do not provide a proof-theoretic interpretation for arbitrary
$\pi$-calculus terms.
Caires and Pfenning~\cite{CairesPfenning10} give a propositions-as-types correspondence between
intuitionistic linear logic and session types, interpreting linear logic propositions as session
types for a restricted $\pi$-calculus, $\pi$DILL.  Of particular importance to this work, they
interpret the multiplicative connectives as prefixing, and the exponentials as replicated processes.

Wadler~\cite{Wadler12} adapts Caires and Pfenning's work to classical linear logic, interpreting
proofs as processes in a restricted $\pi$-calculus, CP.
Additionally, Wadler shows that a core session-typed linear functional language, GV, patterned after
a similar language due to Gay and Vasconcelos~\cite{GayVasconcelos10}, may be translated into
CP. However, GV is less expressive than CP: there are proofs which do not correspond to any GV
program.

Our primary contribution is \hgv (Harmonious GV), a version of GV extended with constructs for
session forwarding, replication, and polymorphism. We identify \hgvpi, the session-typed fragment of
\hgv, and give a type-preserving translation from \hgv to \hgvpi ($\lampi{-}$); this translation
depends crucially on the new constructs of \hgv.  We show that \hgv is sufficient to express all
linear logic proofs by giving type-preserving translations from \hgvpi to CP ($\hgvcp{-}$), and from
CP to \hgvpi ($\cptogv{-}$). Factoring the translation of \hgv into CP through $\lampi{-}$
simplifies the presentation, and illuminates regularities that are not apparent in Wadler's original
translation of GV into CP.  Finally, we show that \hgv, \hgvpi, and CP are all equally expressive.

\section{The \hgv Language}
\label{sect:hgv}

This section describes our session-typed language \hgv, contrasting it with Gay and Vasconcelos's
functional language for asynchronous session types~\cite{GayVasconcelos10}, which we call \last, and
Wadler's GV~\cite{Wadler12}.  In designing \hgv, we have opted for programming convenience over
uniformity, while insisting on a tight correspondence with linear logic.
The session types of \hgv are given by the following grammar:
\begin{equations}
  S & ::= & \gvOutput{T}{S} \mid \gvInput{T}{S} \mid
           \gvPlus{\la_i:S_i}{i} \mid \gvChoice{\la_i:S_i}{i} \mid
           \gvEndOutput \mid \gvEndInput
    \mid X \mid \gvDual{X} \mid
            \gvOutputType{X}{S} \mid \gvInputType{X}{S} \mid
            \gvServer{S} \mid \gvService{S}
\end{equations}Types for input ($\gvInput{T}{S}$), output ($\gvOutput{T}{S}$), selection ($\gvPlus{\la_i:S_i}{i}$)
and choice ($\gvChoice{\la_i:S_i}{i}$) are standard. Like GV, but unlike \last, we distinguish
output ($\gvEndOutput$) and input ($\gvEndInput$) session ends; this matches the situation in linear
logic, where there is no conveniently self-dual proposition to represent the end of a
session. Variables and their duals ($X,\gvDual{X}$) and type input ($\gvInputType{X}{S}$) and output
($\gvOutputType{X}{S}$), permit definition of polymorphic sessions. We include a notion of
replicated sessions, corresponding to exponentials in linear logic: a channel of type
$\gvService{S}$ is a ``service'', providing any number of channels of type $S$; a channel of type
$\gvServer{S}$ is the ``server'' providing such a service.
Each session type $S$ has a dual $\gvDual{S}$ (with the obvious dual for variables $X$):
\[
\bl
  \gvDual{\gvOutput{T}{S}} = \gvInput{T}{\gvDual{S}}
\qquad
  \gvDual{\gvPlus{\la_i:S_i}{i}} = \gvChoice{\la_i:\gvDual{S_i}}{i}
\qquad
  \gvDual{\gvEndOutput} = \gvEndInput
\qquad
  \gvDual{\gvOutputType{X}{S}} = \gvInputType{X}{\gvDual{S}}
\qquad
  \gvDual{\gvService{S}} = \gvServer{\gvDual{S}}
\\
  \gvDual{\gvInput{T}{S}} = \gvOutput{T}{\gvDual{S}}
\qquad
  \gvDual{\gvChoice{\la_i:S_i}{i}} = \gvPlus{\la_i:\gvDual{S_i}}{i}
\qquad
  \gvDual{\gvEndInput} = \gvEndOutput
\qquad
  \gvDual{\gvInputType{X}{S}} = \gvOutputType{X}{\gvDual{S}}
\qquad
  \gvDual{\gvServer{S}} = \gvService{\gvDual{S}}
\\
\el
\]
Note that dualisation leaves input and output types unchanged.  In addition to sessions, \hgv's
types include linear pairs, and linear and unlimited functions:
\[
T,U,V ::= S \mid \gvTimes{T}{U} \mid \gvLinFun{T}{U} \mid \gvUnFun{T}{U}
\]
Every type $T$ is either linear ($\lin{T}$) or unlimited ($\un{T}$); the only unlimited types are
services ($\un{\gvService{S}}$), unlimited functions ($\un{\gvUnFun{T}{U}}$), and end input session
types ($\un{\gvEndInput}$).
In GV, $\gvEndInput$ is linear. We choose to make it unlimited in \hgv because then we can dispense
with GV's explicit $\key{terminate}$ construct while maintaining a strong correspondence with
CP---$\gvEndInput$ corresponds to $\cpBottom$ in CP, for which weakening and contraction are
derivable.

\begin{figure}
Structural rules
\begin{mathpar}
\infr[Id]
{ }
{\gvj{x:T}{x}{T}}

\infr[Weaken]
{\gvj{\Phi}{N}{U} \\ \un{T}}
{\gvj{\Phi,x:T}{N}{U}}

\infr[Contract]
{\gvj{\Phi,x:T,x':T}{N}{U} \\ \un{T}}
{\gvj{\Phi,x:T}{\subst{N}{x}{x'}}{U}}
\end{mathpar}Lambda rules
\begin{mathpar}
\infr[$\lolli$-I]
{\gvj{\Phi,x:T}{N}{U}}
{\gvj{\Phi}{\gvLam{x}{N}}{\gvLinFun{T}{U}}}

\infr[$\lolli$-E]
{\gvj{\Phi}{L}{\gvLinFun{T}{U}} \\ \gvj{\Psi}{M}{T}}
{\gvj{\Phi,\Psi}{\gvApp{L}{M}}{U}}

\infr[$\to$-I]
{\gvj{\Phi}{L}{\gvLinFun{T}{U}} \\ \un{\Phi}}
{\gvj{\Phi}{L}{\gvUnFun{T}{U}}}

\infr[$\to$-E]
{\gvj{\Phi}{L}{\gvUnFun{T}{U}}}
{\gvj{\Phi}{L}{\gvLinFun{T}{U}}}

\infr[$\otimes$-I]
{\gvj{\Phi}{M}{T} \\ \gvj{\Psi}{N}{U}}
{\gvj{\Phi,\Psi}{\gvPair{M}{N}}{\gvTimes{T}{U}}}

\infr[$\otimes$-E]
{\gvj{\Phi}{M}{\gvTimes{T}{U}} \\ \gvj{\Psi,x:T,y:U}{N}{V}}
{\gvj{\Phi,\Psi}{\gvLet{\gvPair{x}{y}}{M}{N}}{V}}
\end{mathpar}Session rules
\begin{mathpar}
\infr[Send]
{\gvj{\Phi}{M}{T} \\ \gvj{\Psi}{N}{\gvOutput{T}{S}}}
{\gvj{\Phi}{\gvSend{M}{N}}{S}}

\infr[Receive]
{\gvj{\Phi}{M}{\gvInput{T}{S}}}
{\gvj{\Phi}{\gvReceive{M}}{\gvTimes{T}{S}}}
\\

\infr[Select]
{\gvj{\Phi}{M}{\gvPlus{\la_i:S_i}{i}}}
{\gvj{\Phi}{\gvSelect{\la_j}{M}}{S_j}}

\infr[Case]
{\gvj{\Phi}{M}{\gvChoice{\la_i:S_i}{i}} \\ \row{\gvj{\Psi,x:S_i}{N_i}{T}}{i}}
{\gvj{\Phi,\Psi}{\gvCase{M}{\row{\la_i(x).N_i}{i}}}{T}}
\\

\infr[Fork]
{\gvj{\Phi,x:S}{M}{\gvEndOutput}}
{\gvj{\Phi}{\gvFork{x}{M}}{\gvDual{S}}}

\infr[Link]
{\gvj{\Phi}{M}{S} \\ \gvj{\Phi}{N}{\gvDual{S}}}
{\gvj{\Phi}{\gvLink{M}{N}}{\gvEndOutput}}

\infr[SendType]
{\gvj{\Phi}{M}{\gvOutputType{X}{S'}}}
{\gvj{\Phi}{\gvSendType{S}{M}}{\subst{S'}{S}{X}}}

\infr[ReceiveType]
{\gvj{\Phi}{M}{\gvInputType{X}{S}} \\ X \notin \FV{\Phi}}
{\gvj{\Phi}{\gvReceiveType{X}{M}}{S}}

\infr[Serve]
{\gvj{\Phi,x : S}{M}{\gvEndOutput} \\ \un{\Phi}}
{\gvj{\Phi}{\gvServe{x}{M}}{\gvDual{\gvServer{S}}}}

\infr[Request]
{\gvj{\Phi}{M}{\gvService{S}}}
{\gvj{\Phi}{\gvRequest{M}}{S}}
\end{mathpar}
\caption{Typing rules for \hgv}
\label{fig:hgv-typing}

\end{figure}

Figure~\ref{fig:hgv-typing} gives the terms and typing rules for \hgv; the first block contains the
structural rules, the second contains the (standard) rules for lambda terms, and the third contains
the session-typed fragment.  The $\key{fork}$ construct provides session initiation, filling the
role of GV's $\key{with}\dots\key{connect}\dots\key{to}\dots$ structure, but without the asymmetry
of the latter. The two are interdefinable, as follows:
\[
\gvFork{x}{M} \equiv \gvWith{x}{M}{x} \qquad \gvWith{x}{M}{N} \equiv \gvLet{x}{\gvFork{x}{M}}{N}
\]
We add a construct $\gvLink{M}{N}$ to implement channel forwarding; this form is provided in neither
GV nor \last, but is necessary to match the expressive power of CP. (Note that while we could define
session forwarding in GV or LAST for any particular session type, it is not possible to do so in a
generic fashion.) We add terms $\gvSendType{S}{M}$ and $\gvReceiveType{X}{M}$ to provide session
polymorphism, and $\gvServe{x}{M}$ and $\gvRequest{M}$ for replicated sessions. Note that, as the
body $M$ of $\gvServe{x}{M}$ may be arbitrarily replicated, it can only refer to the unlimited
portion of the environment.
Channels of type $\gvService{S}$ offer arbitrarily many sessions of type $S$; correspondingly,
channels of type $\gvServer{S}$ must consume arbitrarily many $S$ sessions. The rule for
$\gvServe{x}{M}$ parallels that for $\key{fork}$: it defines the server (which replicates $M$) and
returns the channel by which it may be used (of type $\gvDual{\gvServer{S}} =
\gvService{\gvDual{S}}$). As a consequence, there is no rule involving type $\gvServer{S}$. We
experimented with having such a rule, but found that it was always used immediately inside a
$\key{fork}$, while providing no extra expressive power. Hence we opted for the rule presented here.

\section{From \hgv to \hgvpi}
\label{sect:hgv-to-hgvpi}

The language \hgvpi is the restriction of \hgv to session types, that is, \hgv without $\lolli$,
$\to$, or $\otimes$. In order to avoid $\otimes$, we disallow plain $\gvReceive{M}$, but do permit
it to be fused with a pair elimination $\gvReceiveK{x}{y}{M}{N}$. We can simulate all non-session
types as session types via a translation from \hgv to \hgvpi. The translation on types is given by
the homomorphic extension of the following equations:
\[
\lampi{\gvLinFun{T}{U}} = \gvOutput{\lampi{T}}{\lampi{U}} \qquad
\lampi{\gvUnFun{T}{U}} = \gvService{(\gvOutput{\lampi{T}}{\lampi{U}})} \qquad
\lampi{\gvTimes{T}{U}} = \gvInput{\lampi{T}}{\lampi{U}}
\]Each target type is the \emph{interface} to the simulated source type. A linear function is
simulated by input on a channel; its interface is output on the other end of the channel. An
unlimited function is simulated by a server; its interface is the service on the other end of that
channel. A tensor is simulated by output on a channel; its interface is input on the other end of
that channel.
This duality between implementation and interface explains the flipping of types in Wadler's
original CPS translation from GV to CP.
The translation on terms is given by the homomorphic extension of the following equations:
\begin{equations}
\lampi{\gvLam{x}{M}} &=& \gvFork{z}{\gvLet{\gvPair{x}{z}}{\gvReceive{z}}{\gvLink{\lampi{M}}{z}}} \\
\lampi{\gvApp{L}{M}} &=& \gvSend{\lampi{M}}{\lampi{L}} \\
\lampi{M,N} &=&
  \gvFork{z}
    {\gvLink{(\gvSend{\lampi{M}}{z})}{\lampi{N}}} \\
\lampi{\gvLet{\gvPair{x}{y}}{M}{N}} &=&
    \gvLet{\gvPair{x}{y}}{\gvReceive{\lampi{M}}}{\lampi{N}} \\
\lampi{L : \gvUnFun{T}{U}} &=&
  \gvServe{z}{\gvLink{\lampi{L}}{z}} \\
\lampi{L : \gvLinFun{T}{U}} &=& \gvRequest{\lampi{L}} \\
\lampi{\gvReceive{M}} &=& \lampi{M} \\
\end{equations}Formally, this is a translation on derivations. We write type annotations to indicate $\to$
introduction and elimination. For all other cases, it is unambiguous to give the translation on
plain term syntax. Each introduction form translates to an interface $\gvFork{z}{M}$ of type
$\gvDual{S}$, where $M : \gvEndOutput$ provides the implementation, with $z : S$ bound in $M$.
We can extend the translation on types to a translation on contexts:
\begin{equations}
\lampi{x_1:T_1, \dots, x_n:T_n} &=& x_1:\lampi{T_1}, \dots, x_n:\lampi{T_n} \\
\end{equations}It is straightforward to verify that our translation preserves typing.
\begin{theorem}
If $\gvj{\Phi}{M}{T}$ then $\gvj{\lampi{\Phi}}{\lampi{M}}{\lampi{T}}$.
\end{theorem}

\section{From \hgvpi to CP}
\label{sect:hgvpi-to-cp}

\begin{figure}
\begin{mathpar}
\infr[Ax]
{ }
{\cpj{\cpLink{w}{x}}{w:\cpDual{A},x:A}}

\infr[Cut]
{\cpj{P}{\G,x:A} \\ \cpj{Q}{\D,x:\cpDual{A}}}
{\cpj{\cpCut{x}{P}{Q}}{\G,\D}}

\infr[$\otimes$]
  {\cpj{P}{\G,y:A} \\ \cpj{Q}{\D,x:B}}
  {\cpj{\cpOutput{x}{y}{P}{Q}}{\G,\D,x:\cpTimes{A}{B}}}

\infr[$\bindnasrepma$]
{\cpj{R}{\Theta,y:A,x:B}}
{\cpj{\cpInput{x}{y}{R}}{\Theta,x:\cpPar{A}{B}}}

\infr[$\oplus$]
{\cpj{P}{\G,x:A_i}}
{\cpj{\cpInject{x}{\la_i}{P}}{\G,x:\cpPlus{\la_i:A_i}{i}}}

\infr[$\binampersand$]
{\row{\cpj{Q_i}{\D,x_i:A_i}}{i}}
{\cpj{\cpCase{x}{\row{\la_i.Q_i}{i}}}{\D,x:\cpWith{\la_i:A_i}{i}}}

\infr[$!$]
{\cpj{P}{\cpWhyNot{\G},y:A}}
{\cpj{\cpServe{x}{y}{P}}{\cpWhyNot{\G},x:\cpOfCourse{A}}}

\infr[$?$]
{\cpj{Q}{\D,y:A}}
{\cpj{\cpRequest{x}{y}{Q}}{\D,x:\cpWhyNot{A}}}

\infr[Weaken]
{\cpj{Q}{\D}}
{\cpj{Q}{\D,x:\cpWhyNot{A}}}

\infr[Contract]
{\cpj{Q}{\D,x:\cpWhyNot{A},x':\cpWhyNot{A}}}
{\cpj{\subst{Q}{x}{x'}}{\D,x:\cpWhyNot{A}}}

\infr[$\exists$]
{\cpj{P}{\G,x:\subst{B}{A}{X}}}
{\cpj{\cpSendType{x}{A}{P}}{\G,x:\cpExists{X}{B}}}

\infr[$\forall$]
{\cpj{Q}{\D,x:B} \\ X \notin \D}
{\cpj{\cpReceiveType{x}{X}{Q}}{\D,x:\cpForall{X}{B}}}

\infr[$\cpOne$]
{ }
{\cpj{\cpEmptyOut{x}}{x:\cpOne}}

\infr[$\cpBottom$]
{\cpj{P}{\G}}
{\cpj{\cpEmptyIn{x}{P}}{\G,x:\cpBottom}}
\end{mathpar}
\caption{Typing rules for CP}\label{fig:cp-typing}
\end{figure}

We present the typing rules of CP in Figure~\ref{fig:cp-typing}.  Note that the propositions of CP
are exactly those of classical linear logic, as are the cut rules (if we ignore the terms). Thus, CP
enjoys all of the standard meta theoretic properties of classical linear logic, including confluence
and weak normalisation.
A minor syntactic difference between our presentation and Wadler's is that our sum ($\oplus$) and
choice ($\binampersand$) types are $n$-ary, matching the corresponding session types in \hgv,
whereas he presents binary and nullary versions of sum and choice.
Duality on CP types ($\cpDual{(-)}$) is standard:
\[
\bl
  \cpDual{(\cpTimes{A}{B})} \!=\! \cpPar{\cpDual{A}}{\cpDual{B}}
~~
  \cpDual{(\cpPlus{\la_i:A_i}{i})} \!=\! \cpWith{\la_i:\cpDual{A_i}}{i}
~~
  \cpDual{\cpOne} \!=\! \cpBottom
~~
  \cpDual{(\cpExists{X}{B})} \!=\! \cpForall{X}{\cpDual{B}}
~~
  \cpDual{(\cpOfCourse{A})} \!=\! \cpWhyNot{\cpDual{A}}
\\
  \cpDual{(\cpPar{A}{B})} \!=\! \cpTimes{\cpDual{A}}{\cpDual{B}}
~~
  \cpDual{(\cpWith{\la_i:A_i}{i})} \!=\! \cpPlus{\la_i:\cpDual{A_i}}{i}
~~
  \cpDual{\cpBottom} \!=\! \cpOne
~~
  \cpDual{(\cpForall{X}{B})} \!=\! \cpExists{X}{\cpDual{B}}
~~
  \cpDual{(\cpWhyNot{A})} \!=\! \cpOfCourse{\cpDual{A}}
\\
\el
\]

The semantics of CP terms follows the cut elimination rules in classical linear logic.  We interpret
the cut relation $\redto$ modulo $\alpha$-equivalence and structural cut equivalence:
\begin{equations}
\cpLink{x}{y} &\eqto& \cpLink{y}{x} \\
\cpCut{x}{P}{Q} &\eqto& \cpCut{x}{Q}{P} \\
\cpCut{y}{\cpCut{x}{P}{Q}}{R} &\eqto& \cpCut{x}{P}{\cpCut{y}{Q}{R}} \\
\end{equations}
The principal cut elimination rules correspond to communication between processes.
\begin{equations}
\cpCut{x}{\cpLink{w}{x}}{P}
  &\redto& \subst{P}{w}{x} \\
\cpCut{x}{\cpOutput{x}{y}{P}{Q}}{\cpInput{x}{y}{R}}
  &\redto& \cpCut{y}{P}{\cpCut{x}{Q}{R}} \\
\cpCut{x}{\cpInject{x}{\la_j}{P}}{\cpCase{x}{\row{\la_i.Q_i}{i}}}
  &\redto& \cpCut{x}{P}{Q_j} \\
\cpCut{x}{\cpServe{x}{y}{P}}{\cpRequest{x}{y}{Q}}
  &\redto& \cpCut{y}{P}{Q} \\
\cpCut{x}{\cpServe{x}{y}{P}}{Q}
  &\redto& Q, \quad x \notin \FV{Q} \\
\cpCut{x}{\cpServe{x}{y}{P}}{\subst{Q}{x}{x'}}
  &\redto& \cpCut{x}{\cpServe{x}{y}{P}}{\cpCut{x'}{\cpServe{x'}{y}{P}}{Q}} \\
\cpCut{x}{\cpSendType{x}{A}{P}}{\cpReceiveType{x}{X}{Q}}
  &\redto& \cpCut{x}{P}{\subst{Q}{A}{X}} \\
\cpCut{x}{\cpEmptyOut{x}}{\cpEmptyIn{x}{P}}
  &\redto& P \\
\end{equations}
Finally, we provide commuting conversions, moving communication under unrelated prefixes.
\begin{equations}
\cpCut{z}{\cpOutput{x}{y}{P}{Q}}{R}
  &\redto& \cpOutput{x}{y}{\cpCut{z}{P}{R}}{Q}, \quad z \in \FV{P} \\
\cpCut{z}{\cpOutput{x}{y}{P}{Q}}{R}
  &\redto& \cpOutput{x}{y}{P}{\cpCut{z}{Q}{R}}, \quad z \in \FV{Q} \\
\cpCut{z}{\cpInput{x}{y}{P}}{Q}
  &\redto& \cpInput{x}{y}{\cpCut{z}{P}{Q}} \\
\cpCut{z}{\cpInject{x}{\la}{P}}{Q}
  &\redto& \cpInject{x}{\la}{\cpCut{z}{P}{Q}} \\
\cpCut{z}{\cpCase{x}{\row{\la_i.Q_i}{i}}}{R}
  &\redto& \cpCase{x}{\row{\la_i.\cpCut{z}{Q_i}{R}}{i}} \\
\cpCut{z}{\cpServe{x}{y}{P}}{Q}
  &\redto& \cpServe{x}{y}{\cpCut{z}{P}{Q}} \\
\cpCut{z}{\cpRequest{x}{y}{P}}{Q}
  &\redto& \cpRequest{x}{y}{\cpCut{z}{P}{Q}} \\
\cpCut{z}{\cpSendType{x}{A}{P}}{Q}
  &\redto& \cpSendType{x}{A}{\cpCut{z}{P}{Q}} \\
\cpCut{z}{\cpReceiveType{x}{X}{P}}{Q}
  &\redto& \cpReceiveType{x}{X}{\cpCut{z}{P}{Q}} \\
\cpCut{z}{\cpEmptyIn{x}{P}}{Q}
  &\redto& \cpEmptyIn{x}{\cpCut{z}{P}{Q}} \\
\end{equations}
A fuller account of CP can be found in Wadler's work~\cite{Wadler12}.

We now give a translation from \hgvpi to CP. Post composing this with the embedding of \hgv in
\hgvpi yields a semantics for \hgv.
The translation on session types is as follows:
\[
\ba{@{}c@{\qquad}c@{\qquad}c@{\qquad}c@{}}
\begin{eqs}
\hgvcp{\gvOutput{T}{S}}         &=& \cpTimes{\cpDual{\hgvcp{T}}}{\hgvcp{S}} \\
\hgvcp{\gvInput{T}{S}}          &=& \cpPar{\hgvcp{T}}{\hgvcp{S}} \\
\hgvcp{\gvEndOutput}  &=& \cpOne \\
\end{eqs}
&
\begin{eqs}
\hgvcp{\gvPlus{\la_i:S_i}{i}}   &=& \cpPlus{\la_i:\hgvcp{S_i}}{i} \\
\hgvcp{\gvChoice{\la_i:S_i}{i}} &=& \cpWith{\la_i:\hgvcp{S_i}}{i} \\
\hgvcp{\gvEndInput}   &=& \cpBottom \\
\end{eqs}
&
\begin{eqs}
\hgvcp{\gvServer{S}}  &=& \cpOfCourse{\hgvcp{S}} \\
\hgvcp{\gvService{S}} &=& \cpWhyNot{\hgvcp{S}} \\
\hgvcp{X}                   &=& X \\
\end{eqs}
&
\begin{eqs}
\hgvcp{\gvOutputType{X}{S}} &=& \cpExists{X}{\hgvcp{S}} \\
\hgvcp{\gvInputType{X}{S}}  &=& \cpForall{X}{\hgvcp{S}} \\
\hgvcp{\gvDual{X}}          &=& \cpDual{X} \\
\end{eqs}
\ea
\]
The translation is homomorphic except for output, where the output type is dualised. This accounts
for the discrepancy between $\gvDual{\gvOutput{T}{S}} = \gvInput{T}{\gvDual{S}}$ and
$\cpDual{(\cpTimes{A}{B})} = \cpPar{\cpDual{A}}{\cpDual{B}}$.

The translation on terms is formally specified as a CPS translation on derivations as in Wadler's
presentation. We provide the full translations of weakening and contraction for $\gvEndInput$, as
these steps are implicit in the syntax of \hgv terms. The other constructs depend only on the
immediate syntactic structure, so we abbreviate their translations as mappings on plain terms:
\begin{equations}
\hgvcpl \inferrule{\gvj{\Phi}{N}{S}}{\gvj{\Phi,x:\gvEndInput}{N}{S}} \hgvcpr\!\!z &=&
  \inferrule
    {\cpj{\hgvcp{N}z}{\hgvcp{\Phi},z:\cpDual{\hgvcp{S}}}}
    {\cpj{\cpEmptyIn{x}{\hgvcp{N}z}}{\hgvcp{\Phi},x:\cpBottom,z:\cpDual{\hgvcp{S}}}}
\\[3ex]
\hgvcpl \inferrule{\gvj{\Phi,x:\gvEndInput,x':\gvEndInput}{N}{S}}
                          {\gvj{\Phi,x:\gvEndInput}{\subst{N}{x}{x'}}{S}} \hgvcpr\!\!z &=&
  \inferrule
    {\cpj{\hgvcp{N}z}{\hgvcp{\Phi},x:\cpBottom,x':\cpBottom,z:\cpDual{\hgvcp{S}}}}
    {\cpj{\cpCut{x'}{\hgvcp{N}z}{\cpEmptyOut{x'}}}{\hgvcp{\Phi},x:\cpBottom,z:\cpDual{\hgvcp{S}}}}
\\[3ex]
\hgvcp{x}z &=& \cpLink{x}z \\
\hgvcp{\gvSend{M}{N}}z &=& \cpCut{x}{\cpOutput{x}{y}{\hgvcp{M}y}{\cpLink{x}{z}}}{\hgvcp{N}x} \\
\hgvcp{\gvReceiveK{x}{y}{M}{N}}z &=&
  \cpCut{y}{\hgvcp{M}y}{\cpInput{y}{x}{\hgvcp{N}z}} \\
\hgvcp{\gvSelect{\la}{M}}z &=&
  \cpCut{x}{\hgvcp{M}x}{\cpInject{x}{\la}{\cpLink{x}{z}}} \\
\hgvcp{\gvCase{M}{\row{\la_i(x).N_i}{i}}}z &=&
  \cpCut{x}{\hgvcp{M}x}{\cpCase{x}{\row{\la_i.\hgvcp{N_i}z}{i}}} \\
\hgvcp{\gvFork{x}{M}}z &=&
  \cpCut{x}{\cpCut{y}{\hgvcp{M}y}{\cpEmptyOut{y}}}{\cpLink{x}{z}} \\
\hgvcp{\gvLink{M}{N}}z &=& \cpEmptyIn{z}{\cpCut{x}{\hgvcp{M}x}{\hgvcp{N}x}} \\
\hgvcp{\gvSendType{S}{M}}z &=&
  \cpCut{x}{\hgvcp{M}x}{\cpSendType{x}{\hgvcp{S}}{\cpLink{x}{z}}} \\
\hgvcp{\gvReceiveType{X}{M}}z &=&
  \cpCut{x}{\hgvcp{M}x}{\cpReceiveType{x}{X}{\cpLink{x}{z}}} \\
\hgvcp{\gvServe{y}{M}}z &=&
   \cpServe{z}{y}
      {\cpCut{x}{\hgvcp{M}x}{\cpEmptyOut{x}}} \\
\hgvcp{\gvRequest{M}}z &=& \cpCut{x}{\hgvcp{M}x}{\cpRequest{x}{y}{\cpLink{y}{z}}} \\
\end{equations}Channel $z$ provides a continuation, consuming the output of the process representing the original
\hgvpi term.
The translation on contexts is pointwise.
\begin{equations}
\hgvcp{x_1:T_1, \dots, x_n:T_n} &=& x_1:\hgvcp{T_1}, \dots, x_n:\hgvcp{T_n} \\
\end{equations}As with the translation from \hgv to \hgvpi, we can show that this translation preserves typing.
\begin{theorem}
If $\gvj{\Phi}{M}{S}$ then $\cpj{\hgvcp{M}z}{\hgvcp{\Phi},z:\cpDual{\hgvcp{S}}}$.
\end{theorem}

\section{From CP to \hgvpi}
\label{sect:cp-to-hgvpi}

We now present the translation $\cptogv{-}$ from CP to \hgvpi. The translation on types is as
follows:
\[
\ba{@{}c@{\qquad}c@{\qquad}c@{\qquad}c@{}}
\begin{eqs}
\cptogv{\cpTimes{A}{B}} &=& \gvOutput{\gvDual{\cptogv{A}}}{\cptogv{B}} \\
\cptogv{\cpPar{A}{B}}   &=& \gvInput{\cptogv{A}}{\cptogv{B}} \\
\cptogv{\cpOne}         &=& \gvEndOutput \\
\end{eqs}
&
\begin{eqs}
\cptogv{\cpPlus{\la_i:A_i}{i}} &=& \gvPlus{\la_i:\cptogv{A_i}}{i}  \\
\cptogv{\cpWith{\la_i:A_i}{i}} &=& \gvChoice{\la_i:\cptogv{A_i}}{i} \\
\cptogv{\cpBottom}      &=& \gvEndInput \\
\end{eqs}
&
\begin{eqs}
\cptogv{\cpExists{X}{A}} &=& \gvOutputType{X}{\cptogv{A}} \\
\cptogv{\cpForall{X}{A}} &=& \gvInputType{X}{\cptogv{A}} \\
\cptogv{X}               &=& X \\
\end{eqs}
&
\begin{eqs}
\cptogv{\cpWhyNot{A}}    &=& \gvService{\cptogv{A}} \\
\cptogv{\cpOfCourse{A}}  &=& \gvServer{\cptogv{A}} \\
\cptogv{\cpDual{X}}      &=& \gvDual{X} \\
\end{eqs}
\ea
\]
The translation on terms makes use of $\key{let}$ expressions to simplify the presentation; these
are expanded to \hgvpi as follows:
\[
\gvLet{x}{M}{N} \equiv \lampi{(\lambda x.N) M} \equiv \gvSend{M}{(\gvFork{z}{\gvReceiveK{x}{z}{z}{\gvLink{N}{z}}})}.
\]
\[
\begin{eqs}
  \cptogv{\cpOutput{x}{y}{P}{Q}} &=&
  \gvLet{x}{\gvSend{(\gvFork{y}{\cptogv{P}})}{x}}{\cptogv{Q}} \\
\cptogv{\cpInput{x}{y}{P}} &=&
  \gvLet{\gvPair{y}{x}}{\gvReceive{x}}{\cptogv{P}} \\
\cptogv{\cpInject{x}{\la}{P}} &=&
  \gvLet{x}{\gvSelect{\la}{x}}{\cptogv{P}} \\
\cptogv{\cpCase{x}{\row{\la_i.P_i}{i}}} &=&
  \gvCase{x}{\row{\la_i(x).\cptogv{P_i}}{i}} \\
\cptogv{\cpEmptyOut{x}} &=& x \\
\cptogv{\cpEmptyIn{x}{P}} &=& \cptogv{P} \\
\cptogv{\cpCut{x}{P}{Q}} &=&
  \gvLet{x}{\gvFork{x}{\cptogv{P}}}{\cptogv{Q}} \\
\cptogv{\cpLink{x}{y}} &=& \gvLink{x}{y} \\

\cptogv{\cpSendType{x}{A}{P}} &=&
  \gvLet{x}{\gvSendType{\cptogv{A}}{x}}{\cptogv{P}} \\
\cptogv{\cpReceiveType{x}{X}{P}} &=&
  \gvLet{x}{\gvReceiveType{X}{x}}{\cptogv{P}} \\

\cptogv{\cpServe{s}{x}{P}} &=&
  \gvLink{s}{(\gvServe{x}{\cptogv{P}})} \\
\cptogv{\cpRequest{s}{x}{P}} &=&
  \gvLet{x}{\gvRequest{s}}{\cptogv{P}} \\
\end{eqs}
\]
Again, we can extend the translation on types to a translation on contexts, and show that the
translation preserves typing.
\begin{theorem}
If $\cpj{P}{\G}$ then $\gvj{\cptogv{\G}}{\cptogv{P}}{\gvEndOutput}$.
\end{theorem}

\section{Correctness}
\label{sect:correctness}

If we extend $\hgvcp{-}$ to non-session types, as in Wadler's original presentation
(Figure~\ref{fig:hgvcp-ext}), then it is straightforward to show that this monolithic translation
factors through $\lampi{-}$.
\begin{theorem}
\label{th:factor}
$\hgvcp{\lampi{M}}z \redto^* \hgvcp{M}z$ (where $\redto^*$ is the reflexive transitive closure of $\equiv\redto\equiv$).
\end{theorem}
\noindent
The key soundness property of our translations is that if we translate a term from CP to \hgvpi and
back, then we obtain a term equivalent to the one we started with.
\begin{theorem}
\label{th:soundness}
If $\cpj{P}{\G}$ then $\cpCut{z}{\cpEmptyOut{z}}{\hgvcp{\cptogv{P}}z} \redto^* P$.
\end{theorem}
\noindent
Together, Theorem~\ref{th:factor}~and~\ref{th:soundness} tell us that \hgv, \hgvpi, and CP are
equally expressive, in the sense that every $X$ program can always be translated to an equivalent
$Y$ program, where $X,Y \in \{$\hgv, \hgvpi, CP$\}$.

Here our notion of expressivity is agnostic to the nature of the translations. It is instructive
also to consider Felleisen's more refined notion of expressivity~\cite{Felleisen91}. Both
$\lampi{-}$ and $\cptogv{-}$ are local translations, thus both \hgv and CP are
\emph{macro-expressible}~\cite{Felleisen91} in \hgvpi. However, the need for a global CPS
translation from \hgvpi to CP illustrates that \hgvpi is not macro-expressible in CP; hence \hgvpi
is more expressive, in the Felleisen sense, than CP.

\newcommand{\flip}[1]{\llceil{#1}\rrceil}

\begin{figure}
\[
\ba{@{}cc@{}}
\begin{eqs}
\multicolumn{3}{@{}l@{}}{\textbf{Types}} \\
\multicolumn{3}{@{}l@{}}{\hgvcp{T} = \cpDual{\flip{T}}, T \text{ not a session type}} \\
\multicolumn{3}{@{}l@{}}{\text{where}} \\
\quad\flip{\gvLinFun{T}{U}} &=& \cpPar{\cpDual{\flip{T}}}{\flip{U}} \\
\quad\flip{\gvUnFun{T}{U}} &=& \cpOfCourse{(\cpPar{\cpDual{\flip{T}}}{\flip{U}})} \\
\quad\flip{\gvTimes{T}{U}} &=& \cpTimes{\flip{T}}{\flip{U}} \\
\quad\flip{S} &=& \hgvcp{S} \\
\end{eqs}&
\begin{eqs}
\multicolumn{3}{@{}l@{}}{\textbf{Terms}} \\
\hgvcp{\gvLam{x}{N}}z &=& \cpInput{z}{x}{\hgvcp{N}z} \\
\hgvcp{\gvApp{L}{M}}z &=& \cpCut{y}{\hgvcp{L}y}{\cpOutput{y}{x}{\hgvcp{M}x}{\cpLink{y}{z}}} \\
\hgvcp{L : \gvUnFun{T}{U}}z &=& \cpServe{z}{y}{\hgvcp{L}y} \\
\hgvcp{L : \gvLinFun{T}{U}}z &=& \cpCut{y}{\hgvcp{L}y}{\cpRequest{y}{x}{\cpLink{x}{z}}} \\
\hgvcp{\gvPair{M}{N}}z &=& \cpOutput{z}{y}{\hgvcp{M}y}{\hgvcp{N}z} \\
\hgvcp{\gvLet{\gvPair{x}{y}}{M}{N}}z
  &=& \cpCut{y}{\hgvcp{M}y}{\cpInput{y}{x}{\hgvcp{N}z}} \\
\end{eqs}\ea
\]
\footnotesize {The outer dual appears in the type translation because, as in
  Section~\ref{sect:hgv-to-hgvpi}, we must expose \emph{interfaces} rather than implementations of
  simulated types. As in the definition of $\lampi{-}$ in Section~\ref{sect:hgv-to-hgvpi}, we write
  type annotations to indicate $\to$ introduction and elimination.}

\caption{Extension of $\hgvcp{-}$ to non-session types}
\label{fig:hgvcp-ext}
\end{figure}

\section{Conclusions and Future Work}
\label{sect:conclusion}

We have proposed a session-typed functional language, \hgv, building on similar languages of
Wadler~\cite{Wadler12} and of Gay and Vasconcelos~\cite{GayVasconcelos10}. We have shown that \hgv
is sufficient to encode arbitrary linear logic proofs, completing the correspondence between linear
logic and session types. We have also given an embedding of all of \hgv into its session-typed
fragment, simplifying translation from \hgv to CP.

Dardha et al~\cite{DardhaGS12} offers an alternative foundation for session types through a CPS
translation of $\pi$-calculus with session types into a linear $\pi$-calculus. There appear to be
strong similarities between their CPS translation and ours. We would like to make the correspondence
precise by studying translations between their systems and ours.

In addition we highlight several other areas of future work.  First, the semantics of \hgv is given
only by cut elimination in CP. We would like to give \hgv a semantics directly, in terms of
reductions of configurations of processes, and then prove a formal correspondence with cut
elimination in CP.  Second, replication has limited expressive power compared to recursion; in
particular, it cannot express services whose behaviour changes over time or in response to client
requests.  We believe that the study of fixed points in linear logic provides a mechanism to support
more expressive recursive behaviour without sacrificing the logical interpretation of \hgv.  Finally,
as classical linear logic proofs, and hence CP processes, enjoy confluence, \hgv programs are
deterministic. We hope to identify natural extensions of \hgv that give rise to non-determinism, and
thus allow programs to exhibit more interesting concurrent behaviour, while preserving the
underlying connection to linear logic.

\paragraph{Acknowledgements}
We would like to thank Philip Wadler for his suggestions on the direction of this work, and for his
helpful feedback on the results. This work was funded by EPSRC grant number EP/K034413/1.

\label{sect:bib}
\bibliographystyle{eptcs}
\bibliography{cpgv}

\end{document}